\documentstyle[twoside,multicol,aps,epsf]{revtex}
\begin{document}
\title{Dynamical transitions in the evolution of learning algorithms by selection}

\author{Juan Pablo Neirotti and Nestor Caticha \\
Departamento de F\'{\i}sica Geral, Instituto de F\'{\i}sica, Universidade
de S\~ao Paulo, \emph{}Rua do Mat\~ao Travessa R 187, CEP 05508-900 S\~ao Paulo, Brazil
}
\date{\today}
\maketitle

\vspace{.5cm}
\begin{center}
{\bf Abstract}
\end{center}

\begin{abstract}
We study the evolution of artificial learning systems by means of selection.
Genetic programming is used to generate a sequence of populations of algorithms
which can be used by neural networks for supervised learning of a rule that
generates examples. In opposition to concentrating on final results, which would
be the natural aim while designing good learning algorithms, we study the evolution
process and pay particular attention to the temporal order of appearance of
functional structures responsible for the improvements in the learning process,
as measured by the generalization capabilities of the resulting algorithms.
The effect of such appearances can be described as dynamical phase transitions.
The concepts of phenotypic and genotypic entropies, which serve to describe
the distribution of fitness in the population and the distribution of symbols
respectively, are used to monitor the dynamics. In different runs the phase
transitions might be present or not, with the system finding out good solutions,
or staying in poor regions of algorithm space. Whenever phase transitions occur,
the sequence of appearances are the same. We identify combinations of variables
and operators which are useful in measuring experience or performance in rule
extraction and can thus implement useful annealing of the learning schedule.
We also find combinations that can signal surprise, measured, on a single example,
by the difference between prediction and the correct output. Structures that
measure performance always appear after those for measuring surprise. Invasions
of the population by such structures in the reverse order were never observed.

\end{abstract}

\begin{multicols}{2}

\section{Introduction}

In this paper we consider the dynamics of automatic design of learning algorithms
for neural networks. We use Genetic Programming (GP) as a tool to generate a
sequence of generations of populations of programs which implement a learning
algorithm. Programs at one generation give rise through cross over and mutations
to offspring programs in the next generation according to their fitness. The
fitness -which defines the problem- is related in this study to the efficiency
of the learning algorithm implemented by the program. We choose a measure of
efficiency based on the ability of generalization, related to the expected error
of the output on examples which are statistically independent from the training
set. 

Although GP is similar in spirit and actually inspired by the Genetic Algorithm
(GA) of Holland \cite{Holland}, since both mimic natural evolution, the idea
of GP put forward by Koza \cite{Koza}, differs from GA in very important ways.\footnote{%
In brief, GA optimizes a function which depends on a parameter vector by studying
the evolution of a population of such vectors or strings of parameter values,
generating offspring vectors by several operations, such as cross-over, mutations,
etc.
} GP deals with programs, represented by strings of symbols -variables or functional
operators- which can have as inputs different types of variables and operators
across the population, as well as along the generations. In some loose sense
GP allows for great variability and thus for the emergence of more of something
that could be dubbed complexity. 

Usually the automatic design of programs has as an aim the solution of a problem
and a measure of how far a given candidate goes in that direction is given by
the fitness. We are not only interested in final results, but rather the road
towards that goal and its characterization are the main issues. We have chosen
to study perceptron learning, a sufficiently simple learning problem that can
be studied, as far as final results are concerned, by analytical means but which
presents a wealth of interesting results. By analyzing the development of learning
algorithms we expect to learn something about the dynamics along which different
variable combinations become useful and invade the population of programs. We
find dynamic phase transitions as different functional structures change from
being irrelevant to useful and find evidence that point to a strict temporal
order in the sequence of such appearances. Some structures, though useful at
later stages are irrelevant at first and remain so until some other structure
is mature enough and thus potentialize the utility of the former. 

The characterization of a population can be made through the use of several
different complementary tools. What we call the phenotypic or functional level
description deals with quantities that measure the expression of important traits.
At this level the program differences are irrelevant as long as they give rise
to the implementations of the same function. The main tool, the phenotypic entropy
\( S \) describes the distribution of fitness in the population. At the genotypic
or program level, different programs are different even if they give rise to
the same numbers, for their potential of generating new successful programs
in the following generation depends on the particular symbols which exist at
present. We can introduce several genotypic entropies which describe the distribution
of probabilities of symbols, of two contiguous symbols and so on. We will restrict
to dealing with single symbol distributions and we characterize them by \( H \)
the genotypic entropy \cite{Adami}. 

The crossover of programs, obtained by a cutting and pasting process described
bellow, can be difficult to implement in common programming languages such as
C and Fortran. The major part of the work in GP has been developed in LISP which
is also the case in this study. We have developed also a protocol for simulation
of LISP on a parallel architecture on a cluster of machines running Linux, which
is described in \cite{NeiCat3}. 

The paper is organized as follows. In section 2 a brief description of GP from
the very special point of view which interests us here is followed by a description
of the problem which GP aims at solving. Section 3 presents the results and
concluding remarks can be found in the last section.

\section{The problem and the method}

\subsection{Problem: Learning by a perceptron}

The learning problem to be analyzed by the GP must strike a balance between
being complex enough so that interesting dynamics arises and simple to the point
that details can be understood and simulations performed. The perceptron meets
these demands and has a long and distinguished history. For an extensive view
from a Statistical Mechanics perspective see \cite{AndreasChris}. We consider
the realizable teacher-student learning scenario. The perceptron classifies
vectors \( {\bf S}\in {\rm I}\! {\rm R}^{N} \) (here obtained i.i.d from
a uniform distribution) in two categories with labels \( \sigma _{{\bf J}\mu }=\pm 1 \)
according to the rule \( \sigma _{{\bf J}\mu }={\rm sgn}({\bf J}\cdot {\bf S}) \).
The objective of the learning dynamics is to determine the weight or synaptic
vector \( {\bf J}\in {\rm I}\! {\rm R}^{N} \) from pairs of examples \( ({\bf S}_{\mu },\sigma _{\mathbf{B}\mu }) \)
which carry information about a rule. We restrict ourselves to the simplest
case of noiseless realizable rules, which mean that the labels were uncorrupted
and generated by another perceptron with a weight vector \( {\bf B}\in {\rm I}\! {\rm R}^{N} \)
unknown to us. We consider on-line learning, which means that \( {\bf J} \)
will be built sequentially by modifications induced by the arrival of new pairs
of examples. We even concentrate on the particular form of modulated Hebbian
learning, where the increments of \( {\bf J} \) are described by a modulation
function \( f \), thus \( \Delta {\bf J}_{\mu }={\bf J}_{\mu }-{\bf J}_{\mu -1}=f\, \sigma _{\mathbf{B}\mu }{\bf S}_{\mu }/\sqrt{N} \).
This is not very restrictive as a large fraction of the previously studied algorithms,
both on-line and off-line, may be put in a similar way and in the (thermodynamic)
limit of large networks it can represent asymptotically efficient learning,
which even saturate Bayesian bounds. 

We deal with questions about the modulation function, such as : (i) What are
the variables upon which the modulation function depends? (ii) What is the best
function? (iii) In the event that the machine has no access to all of the useful
variables, which ones can be left out and which are relevant? That is, in the
path towards the development of a more sophisticated algorithm, the machines
at earlier stages may not dispose all relevant variables, then which are the
ones that are relevant in the earlier stages and which become so later? Is there
any discernible pattern in the order these variables are incorporated? That
we can indeed identify such time ordering in our simulations is the main result
of this paper.

Related questions have been addressed before \cite{Amari}, see: about best
results and Bayesian bounds \cite{Opperhau}, for a variational point of view
about the perceptron learning in \cite{KiCa}, about feedforward architectures
with hidden units in \cite{CoCa,ViCa,SiCa}, for drifting rules in \cite{BiehlSchw,KiCa93},
in an unsupervised scenario \cite{VdBReiman}, from a more general Bayesian
perspective in \cite{OpperPRL,Opper,SaraOle}; in the case of off-line learning
in \cite{Opperhau,KiCaOff}. From the perspective of time ordering it has been
discussed in \cite{KiCaTempOrd}.

\subsection{Method: Algorithm construction by Genetic Programming}

In this section we describe briefly our implementation of GP for the problem
at hand. We do not consider the evolution of machine architecture, which is
left for future work and just deal with the evolution of the modulation function. 

Conventional GA work manipulating fixed-length character strings that represent
candidate solutions of a given problem. For many problems, hierarchical computer
programs are the most natural representation for the solution. Since the size
and the shape of the program that represents the solution are unknown in advance,
the program should have the potential of changing its size and shape. The aim
of GP is getting computers to program themselves by providing a domain independent
way to search the space of possible computer programs, for one that solves a
given problem. The principle that rules GP is, as in GA, the survival of the
fittest.

Starting from a population of randomly created computer programs, the GP operations
are used to generate the population of the next generation. The programs are
ranked by their fitness and then the GP operations are applied again. These
two steps are then iterated. 

The most common computer language used in GP is LISP, therefore we will refer
to the population individuals as programs or LISP S-expressions indistinctly.
We call \emph{Faithful S-expressions} (FSEs) a lists of symbols that do not return
an error message when evaluated. Components, also called atoms, of the S-expressions
can be either functional operators or variables. The set of all operators used
in the S-expressions is $ {\cal F} $ and the set of all variables is ${\cal V }$.
The choice of these sets depends on the nature of the problem being faced. For
instance, if the solution of a problem can be represented by a quotient of polynomials,
${\cal F}$ = {\sf \{+ - * /\}} and ${\cal V}$ = {\sf \{x 1\}}. For example,
a FSE is {\sf (+ (+ x x) (* x (- x (- x x))))}, which is a (non unique) LISP
representation of the function \( 2x+x^{2} \). The simplest FSE is an operator
followed by the appropriate number of variables (two in the example above).
All FSEs have an operator as first element, and following elements should be
variables or FSEs. Unfaithful S-expressions are for instance: {\sf (x x)}, {\sf (+ x *)}
and {\sf (x - x)}.

LISP's most prominent characteristic with regard to GP is that programs
and data have a common form and are treated in the same manner. This common
form is equivalent to the parse tree for the computer program and allows to
genetically manipulate the parts of the program (i.e., subtrees of the parse
tree).

\begin{figure}
\epsfxsize=.48\textwidth
\epsfbox{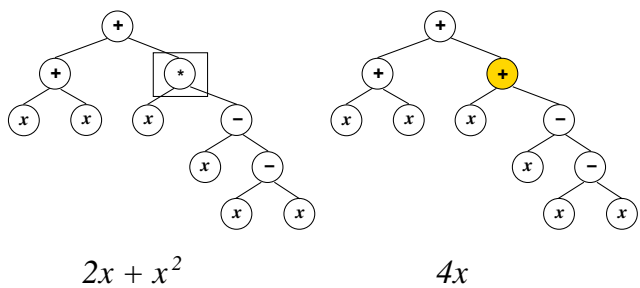}
\caption{LISP programs as parsing trees before and after a GP mutation. A randomly selected
atom in the parse tree is changed to another randomly selected atom of the same
type. In this example a multiplication operation is replaced by an addition.}
\end{figure}

The GP operations considered in the present work are asexual reproduction, mutation
and cross-over. In the operation of asexual reproduction a certain fraction
of the top ranked individuals are copied without any modification into the new
generation, ensuring the preservation of structures that made them successful.
Mutation is implemented by randomly changing an atom of an individual chosen
at random. The new and old atoms must be of the same kind to ensure faithfulness.
Finally the modified tree is copied into the new generation. In order to accelerate
the dynamics different mutation rates can be used for different atom types.
Although there are no sexes associated to the programs, cross-over can be better
described as the sexual GP operation. In our experiments, the first parent is
chosen among the reproduced fraction of the population (those programs that
have been copied from the past generation) by tournament \cite{Koza}. The second
parent is chosen by tournament among the entire population. An atom is selected
randomly in each parent. The subtrees (or leaves) with roots in the selected
atoms are interchanged to generate two offsprings. In order to avoid uncontrolled
growth if the depth of any of the offsprings is above a given threshold, the
program is deleted.

After a new population is created, the fitness of each individual is measured
and so a new ranking is built. There is a great freedom in choosing the fitness
function. It is always a macroscopic or phenotypic quantity, i.e. a function
of the expressed characters, and although it reflects the microstructure, it
is not a function of the genetic details of the individual. Errors in the measurement
of the fitness have a bearing on the dynamics, not entirely different from the
temperature in simulated annealing. 

Our numerical experiments have been performed in a Pentium III, 800 MHz PC,
Linux cluster, using the strategy described in \cite{NeiCat3}. The GP parameters
used in the simulation are presented in Table 1.
At generation zero a population of 500 faithful S-expressions is created at
random. The programs have (in agreement with Table 1) a maximum
depth of \( 7 \) nested parenthesis. The sets used to build the programs are 
\[
{\cal V}=\left\{ \sigma _{{\bf J}\mu }\; \sigma _{\mathbf{B}\mu }\; h\; {\bf S}_{\mu }\; {\bf J}_{\mu }\right\} \, ,\]
where \( h={\bf S}_{\mu }\cdot {\bf J}_{\mu }/\left\Vert {\bf J}_{\mu }\right\Vert  \),
and
\[
{\cal F}=\left\{ {\sf Psqr\; Pexp\; Plog\; abs\; +\; -\; *\; \%\; p.\; pN.\; ev\! *\; vv\! +\; vv\! -}\right\} \, ,\]
 where {\sf Psqr}, {\sf Pexp}, {\sf Plog}, and \% are the protected square
root, exponential, logarithm and division; {\sf abs}, {\sf +}, {\sf -}, and {\sf *} are
the usual absolute value, addition, subtraction and multiplication; and {\sf p.},
{\sf pN.}, {\sf ev\( * \)}, {\sf vv\( + \)}, and {\sf vv\( - \)} are the
inner product, normalized inner product, the product of a scalar times a vector,
the addition of two vectors and the subtraction of two vectors respectively.
Protected functions are functions whose definition domains have been extended
in order to accept a larger set of arguments. The definitions of these functions
appear in Table 2.

\begin{figure}
\epsfxsize=.48\textwidth
\epsfbox{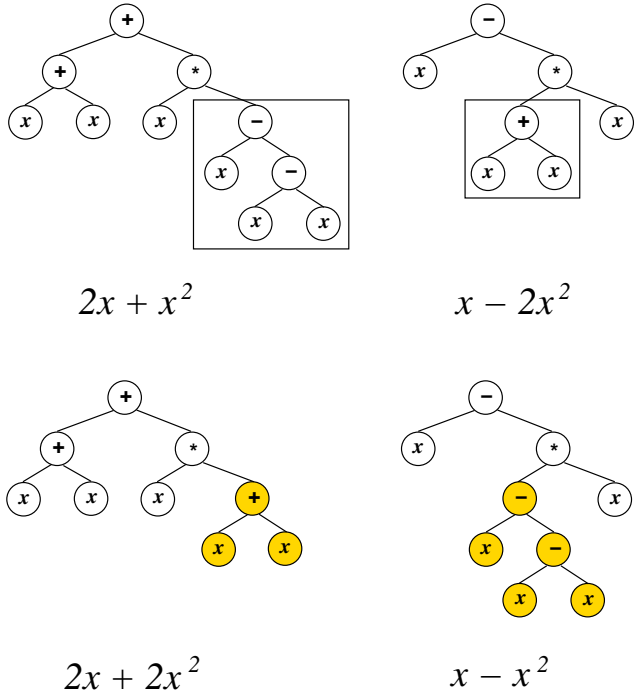}
\caption{GP cross-over. Two parents are selected from the population. A random point
in each tree is selected. The branches that grow from the point are interchanged
in order to generate two offsprings.}
\end{figure}

The inner product is the usual inner product among vectors. If \( {\bf x} \),
\textbf{}\( {\bf y} \) \textbf{\( \in {\rm I}\! {\rm R}^{N} \)} then the
normalized inner product is \( {\bf x}\cdot {\bf y}/N \). Other operations
involving vectors have to be defined. The \( {\sf ev*} \) takes two arguments,
a scalar {\sf x} and a vector \( \mathbf{v} \), and returns a vector \( \mathbf{w} \)
\textbf{}with components \( {\sf w}_{i}={\sf x\, v}_{i} \). The sum (difference)
of two vectors \( {\sf vv+} \) (\( \sf vv- \)) takes two vectors \textbf{\( \mathbf{v} \)}
and \( \mathbf{w} \), and returns a vector \( \mathbf{z} \), with components
\( {\sf z}_{i}={\sf v}_{i}{\sf +w}_{i} \) \( \,  \) \( \left( {\sf z}_{i}={\sf v}_{i}{\sf -w}_{i}\right)  \).

When the process of creation of programs is done, before performing the GP operations
to generate the next generation, the fitness has to be calculated. Because the
programs represent the learning algorithm of a neural network, a good measure
of the learning algorithm performance, should be based on the generalization
error, which measures the probability that the classification of the network
\( \sigma _{{\bf J}\mu } \) is different from the correct label \( \sigma _{\mathbf{B}\mu } \)
\[
e_{g}\left( \mu \right) =\left\langle \Theta \left( -\sigma _{\mathbf{B}\mu }\sigma _{{\bf J}\mu }\right) \right\rangle _{{\cal L}_{\mu }},\]
which in the thermodynamic limit
\[
e_{g}\left( \mu \right) =\frac{1}{\pi }\arccos \rho \, ,\]
where \( \rho =\lim _{N\rightarrow \infty }\left( \frac{{\bf J}_{\mu }\cdot \mathbf{B}}{\left\Vert {\bf J}_{\mu }\right\Vert \left\Vert \mathbf{B}\right\Vert }\right)  \)
and \( \mu  \) indicates how many examples have been presented to the network,
which we call the age of the individual. The average is over training sets of
\( \mu  \) pairs of examples. Since the aim is to obtain algorithms with the
smallest possible generalization error, which depends on the age -taken as the
number of examples already to which the network has been exposed- , we chose
a fitness that incorporates the variation of \( e_{g} \) with age and average
over age so that the asymptotic stage  is at least as important as earlier stages.
For the \( k^{th} \) member of the population
\[
F^{(k)}=\sum ^{P}_{\mu =1}\mu e_{g}\left( \mu \right) \, \]
is the fitness and \( P \) is the total number of examples (maximum age) presented
to the network. The population is ranked according to fitness and the best \( 10\% \)
are asexually reproduced into the next generation (according to the Reproduction
Rate on Table 1). The other \( 90\% \) is generated by cross-over.
The first parent is chosen from the best \( 10\% \) of the population. To do
so we select first a number \( a \) such that \( 1\leq a\leq P \) with a probability
proportional to \emph{\( a \)} (the higher the \emph{\( a \)} the higher the
probability to choose it). \emph{\( a \)} is the age of the individuals that
are going to participate in the tournament. From the best \( 10\% \) of the
population, ten individuals are selected at random. From comparison of their
generalization error at age \emph{\( a \)}, the individual with smaller \( e_{g}(a) \)
is selected for cross-over. To select the second parent a similar mechanism
is applied. Ten individuals are selected at random from the entire population,
and their generalization errors at age \emph{\( a \)} are compared. The winner
is chosen to mate. To perform the cross-over, sub-trees of both parents are
selected at random. Internal points (i.e. operators) are selected more frequently
than external points (i.e. variables) in order to make the individuals grow
(see Table 1).\footnote{%
Observe that an operator in a S-expression is always a root of another S-expression,
while a variable is a S-expression by itself.
}

If either one of the offsprings has a depth bigger than 17, it is deleted. With
a mutation rate of 0.01\% (one every 20 generations) a mutation is performed
to the offsprings. Because the pairs \( {\bf J}_{\mu }\, {\bf J}_{\mu } \)
become rare after few generations (at the beginning of the simulation, the learning
algorithms that use \textbf{J} are not efficient) we keep injecting this pair
with a rate of 0.2\% (at least one individual per generation receives this pair).
Different mutation rates just serve the purpose of accelerating the dynamics
and decrease the time scale of the typical time that it takes for interesting
things to happen. The process is repeated until the new population reaches the
full size fixed here at 500. To calculate the generalization error an average
is taken over at least 50 sets of examples \( {\cal L}_{P}=\left\{ \left( S_{\mu },\sigma _{\mu }\right) \right\} ^{P}_{\mu =1} \). 

\vspace{.2cm}
\begin{center}\begin{tabular}{|c|c|}
\hline 
\hline 
Parameters&
Values\\
\hline 
Population Size&
500\\
\hline 
Reproduction Rate&
10\%\\
\hline 
Mutation Rate&
0.01\%\\
\hline 
\textbf{JJ} Mutation Rate&
0.2\%\\
\hline 
Max. Depth Gen. 0&
7\\
\hline 
Max. Depth Gen. \emph{G}&
17\\
\hline 
Prob. Internal Point Select. (cross-over)&
90\%\\
\hline 
Tournament Participants&
10\\
\hline 
Vector Sizes&
11\\
\hline 
Maximum Number of Training Examples&
100\\
\hline 
Maximum Number of Sets of Examples&
50\\
\hline 
Slave Processors&
10\\
\hline 
\( \beta  \)&
1\\
\hline 
\hline 
\end{tabular}
\end{center}
\vspace{.3cm}

{\small TABLE 1. Control parameters for the GP simulation in our experiments.}

\vspace{.2cm}
\begin{center}\begin{tabular}{|c|c|}
\hline 
\hline 
Function&
Definition\\
\hline 
{\sf (Psqr x)}&
{\sf (sqrt (abs x))}\\
\hline 
{\sf (Pexp x)}&
{\sf (exp (min 13.0 x))}\\
\hline 
{\sf (Plog x)}&
{\sf (log (max 1.d-17 x))}\\
\hline 
{\sf ({\rm \%} x y)}&
{\sf (if ($>$ 1.d-17 (abs y)) 1.d17 (/ x y))}\\
\hline 
\hline 
\end{tabular}
\end{center}
\vspace{.3cm}

{\small TABLE 2. Definition of the protected function as FSEs. The protected square root is
just the square root of the absolute value of its argument. In this manner we
extended its domain into the negatives. The exponential is well defined in the
reals. Although, in order to avoid overflows we have to impose a cut-off. The
protected logarithm has a cut-off at a small positive number to extend its domain
to the non-positive numbers. And the protected quotient allows the division
by zero (if the absolute value of the denominator is smaller than a tiny number
the protected quotient returns a big number, if not it just returns the usual
quotient).}

\section{Results}

To characterize the distribution of the fitness across the population we introduced
the normalized fitness, a measure of the fraction of the total (exponential)
fitness that an individual has, in a way analogous to the canonical state at
temperature \( \beta  \) (although the system is not in equilibrium with any
temperature reservoir): 
\[
n_{\beta }^{(i)}=\frac{\exp \left( -\beta F^{(i)}\right) }{\sum ^{M}_{k=1}\exp \left( -\beta F^{(k)}\right) }\, ,\]
where \( F^{(k)} \) is the fitness measure of th \( k^{th} \) individual of
the population. Note that smaller values of the fitness are associated to better
performances. The use of the exponential amplifies the importance of the individuals
with better performance and \( \beta  \) was kept equal to \( 1 \). We introduce
the entropy of the normalized fitness
\end{multicols}
\begin{figure}
\epsfxsize=.98\textwidth
\epsfbox{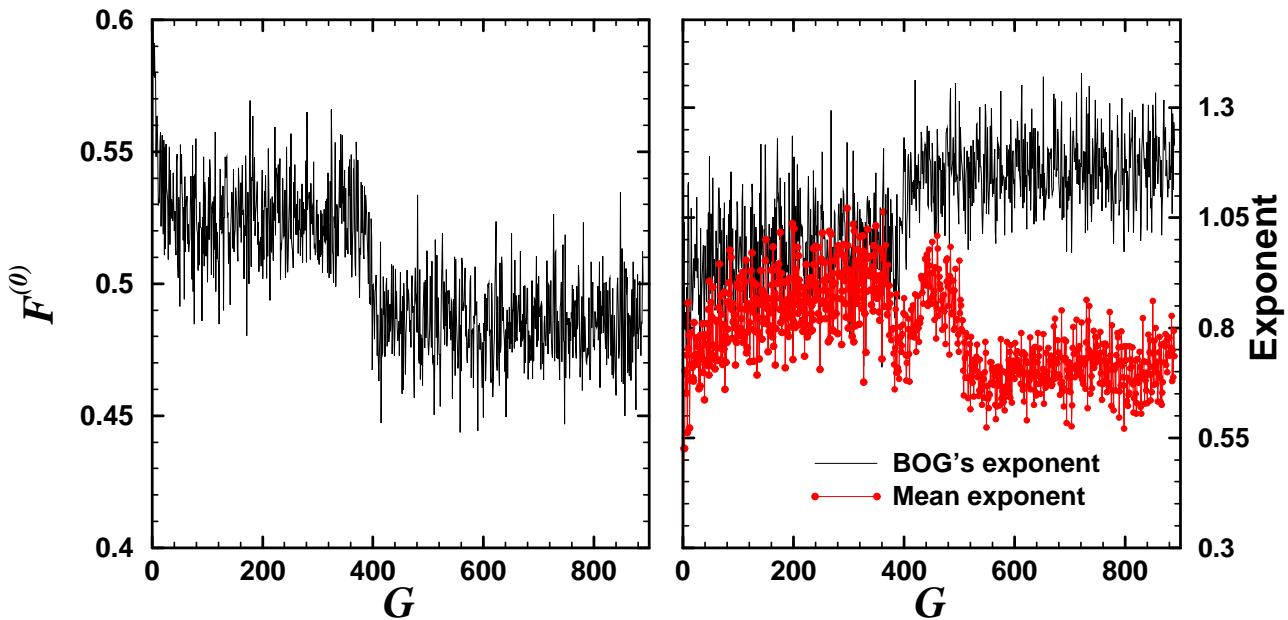}
\caption{(Left) Fitness of the best-of-generation individual of the population vs. the
number of generations. A sudden change takes place around 380 generations. (Right)
The exponent of algebraic decay of \protect\( e_{g}\protect \). Upper curves:
BOG, lower curves: population average.}
\label{fg}
\end{figure}\begin{multicols}{2}
\[
S=-\sum ^{M}_{k=1}n_{\beta }^{(k)}\, \ln \left( n_{\beta }^{(k)}\right) \, ,\]
 a function of the expressed characters of the population (fitness), thus dubbed
the phenotypic entropy or Ph-entropy. Note that this entropy is largest when
all the members of a population have the same fitness and that the appearance
of a distinguished individual, for better or worst, is signaled by a decrease
in Ph-entropy.

Each FSE in the population has a well defined length \( \lambda ^{(k)} \),
i.e the number of atoms (operators and variables) that make it up. We define
the mean length \emph{\( L \)} as
\[
L=\frac{1}{M}\sum _{k=1}^{M}\lambda ^{(k)}\, .\]

To characterize the internal structure of the programs, we estimate for each
position \( i \) the probability that symbol \( s_{q} \) (a variable or an
operator) appears at position \( i \), \( \omega \left( s_{q}|i\right)  \)
by measuring the frequency over all the population. The genotypic entropy (or
G-entropy) which is a function of the micro structure of the individuals in
the population is then defined as \cite{Adami}
\[
H=-\sum _{s_{q}\in {\cal Q}}\sum _{i}\omega \left( s_{q}|i\right) \, \log _{\left| {\cal Q}\right| }\omega \left( s_{q}|i\right) \, ,\]
where \( {\cal Q} \)=${\cal F}$ \( \cup  \) ${\cal V}$

Several numerical experiments, starting from different random seeds, have been
performed using the GP described above. 
\begin{figure}
\epsfxsize=.48\textwidth
\epsfbox{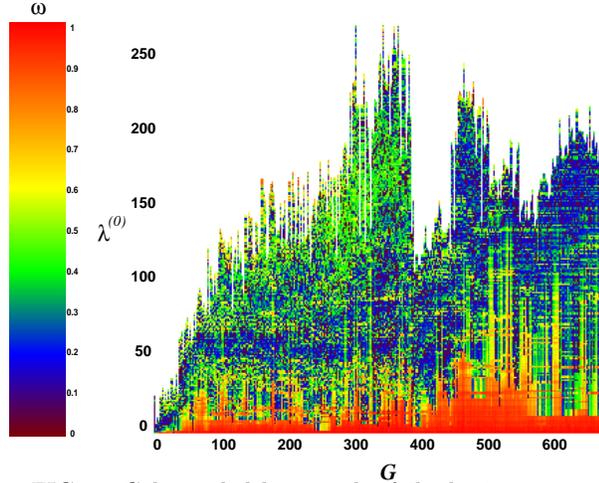}
\caption{Color coded bar graph of the best-in-generation (BOG) individual. The time
\protect\( G\protect \) is in the horizontal axis is measured in generations
and the length of the program is in the vertical axis. Color of pixel at coordinates
\protect\( (G,i)\protect \) codes for the frequency \protect\( \omega \left( s_{q}|i\right) \protect \),
according to the color scale, at which the symbol \protect\( s_{q}\protect \)
(which is the \emph{i}-th atom of the best-of-generation FSE) appears at position
\emph{\protect\( i\protect \),} at generation \emph{\protect\( G\protect \).}}
\label{gen0}
\end{figure}

\noindent Although the history of the population
varies from run to run, we have identified some systematic occurrences. 
\begin{figure}
\epsfxsize=.48\textwidth
\epsfbox{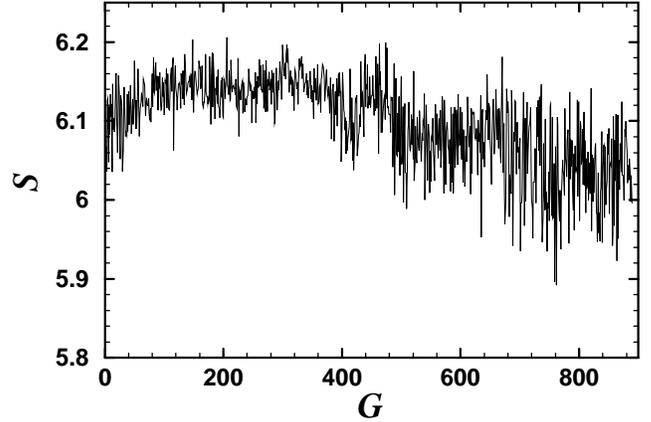}
\caption{Phenotypic entropy as a function of the number of generations 
\protect\( G\protect \).}
\label{entropf}
\end{figure}

\noindent In most
of the runs we have found a drastic change in behavior which can be well described
as a phase transition, although we have neither taken thermodynamic limits associated
to infinite network dimension nor infinite population. The time of the occurrence
varied widely from one simulation to other. In some simulations, the population
did not undergo the transition but it could well happen that we just did not
wait long enough. In what follows we consider an illustrative run which presents
clearly some features that are typical of other runs. We found a dramatic change
of behavior around generation \( 380 \) that can be seen by using several different
signatures. Figure \ref{fg} (left) shows the fitness of the most adapted program
or best-of-generation (BOG) as a function of time. 
The exponent that governs
the decay of \( e_{g} \) shows a sharp change, specially if the population
average is compared to that of the BOG. Finite size errors are responsible for
the fact that exponents larger than one can be found. 
To understand how representative of the whole population is the BOG we composed
a color coded bar graph (see fig. \ref{gen0}) where each vertical bar represents
the BOG program written as a string of symbols, time is measured in generations
in the horizontal axis. At the position of each symbol in the program a colored
square represents the empiric probability of the symbol in the population. Note
that quite rapidly an initial symbol is predominant in the population. This
is invariantly found in all runs and it is always a symbol that ensures that
the modulation function is positive, for other wise the learning would be anti-Hebbian
and inefficient. The initial part of the code is very robust and thus is shared
by almost all the population. 
  There is an obvious change in the length of the BOG which will be considered
bellow, but notice before the transition the upper part is moderately common
(green) and after the transition the upper part is more variable or less frequent.
These changes can also be monitored by the entropies. 

\end{multicols}\begin{figure}
\epsfxsize=.98\textwidth
\epsfbox{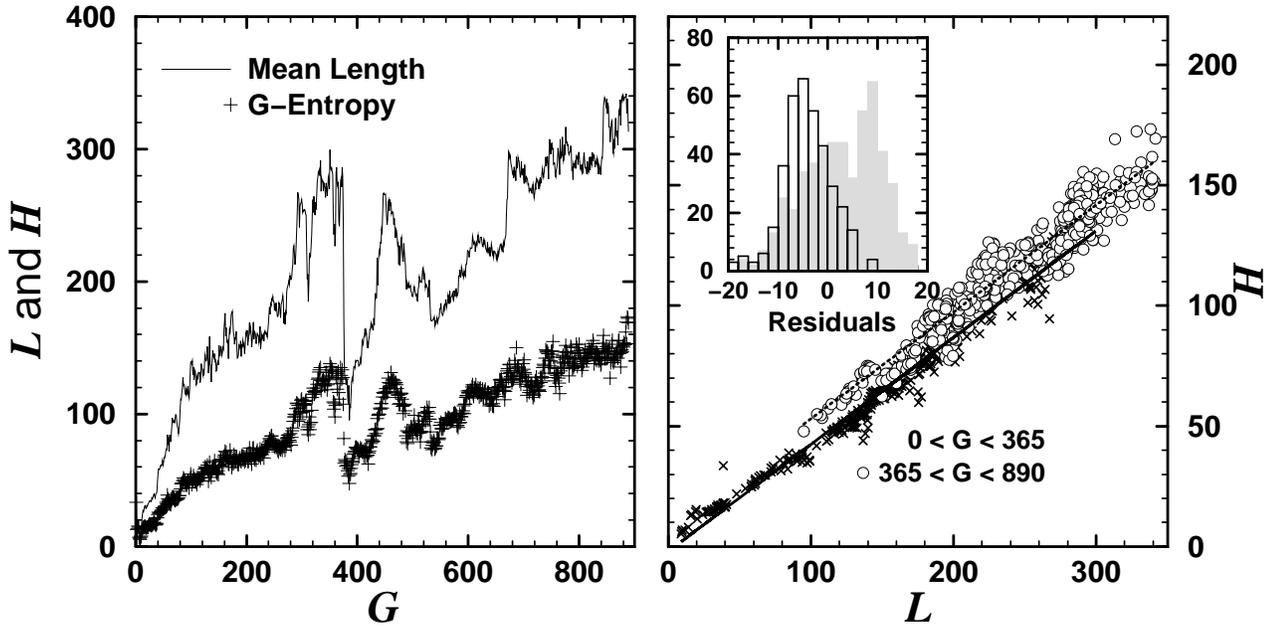}
\caption{(Left) Genotypic entropy and mean length as functions of the number of generations
\protect\( G\protect \). The transition can be seen by the sharp change around
\protect\( G=380\protect \). (Right) \protect\( H\protect \) vs. \protect\( L\protect \).
Two linear fits are shown for data before the transition (crosses) and data
after the transition (circles). To see that two linear fits are necessary we
did a single linear fit of the whole data set and plotted (inset) the histograms
of the residuals to the single linear model. The two histograms are for data
before and after the transition respectively and the separation of the two peaks
lends support to the modeling by two linear regimes.}
\label{entropg}
\end{figure}\begin{multicols}{2}

\begin{figure}
\epsfxsize=.48\textwidth
\epsfbox{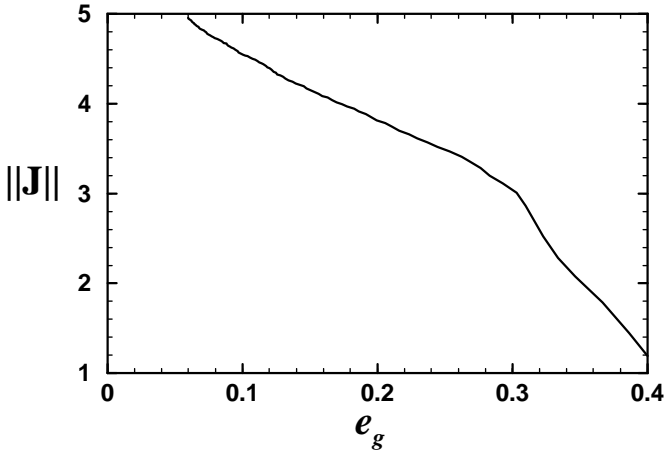}
\caption{Typical behavior for late stage generation: the length of the weight vector
\protect\( ||{\bf J}||\protect \) increases monotonically when the error of
generalization decreases, thus it can be used as a measure of the experience
of the individual or of its performance in solving the classification problem.
It leads to efficient annealing of the learning rates.}
\label{errorj}
\end{figure}

Both entropies (phenotypic and genotypic) present changes about the same time
(figs. \ref{entropf} and \ref{entropg}). The Ph-entropy shows a much larger
variability after the transition, the G-entropy and the mean length both have
an almost discontinuous break at the transition. The fact that the Ph-entropy
has a decreasing trend after the transition can be attributed to the fact that
the G-entropy increases and thereby makes the BOG less frequent than immediately
after the transition. Now the fitness distribution is sharper around the BOG
and therefore the Ph-entropy decreases and oscillates over a wider range. 

The G-entropy and the mean length are linearly correlated. This is natural since
G-entropy, as defined should be extensive. What is not as expected is the fact
that there are two distinct linear regimes before and after the transition.
To see this we did a linear fit to the whole data set and plotted an histogram
of the residuals, that is the difference between the actual value of a data
point and the corresponding value of the linear model. The two histograms in
the inset of fig. \ref{entropg} show clearly a systematic error for the single
linear model. These results prove the existence of a quite sharp transition,
but do not hint at the nature of the changes in the individual programs nor
the reasons for the improvements in fitness. The question is trying to understand
what happened from a functional point of view that led to such an improvement
in generalization ability.

\begin{figure}
\epsfxsize=.48\textwidth
\epsfbox{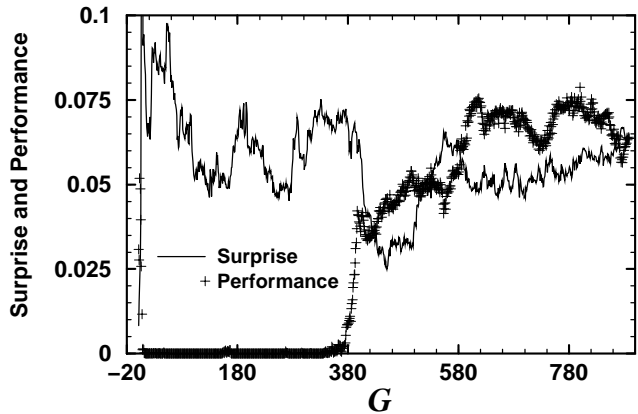}
\caption{Typical behavior of the density of pairs \protect\( h\sigma _{\mathbf{B}\mu }\protect \)
and \protect\( {\bf J}_{\mu }\, {\bf J}_{\mu }\protect \) as functions
of time \protect\( G\protect \) measured in generations. Notice
the sharp rise at the beginning of the density of \protect\( h\sigma _{\mathbf{B}\mu }\protect \)
surprise measuring pairs and the later rise of \protect\( {\bf J}_{\mu }\, {\bf J}_{\mu }\protect \)
, performance measuring pairs. The time ordering is robust and was never seen
in the reverse order.}
\label{syeg}
\end{figure}

There are two quantities or functional structures that are of interest both
in a quantitative and qualitative analysis of the of learning algorithms. The
first, which can be associated to the product \( h\sigma _{\mathbf{B}\mu } \)
can be functionally described as quantifying a measure of surprise. This is
because if \( h\sigma _{\mathbf{B}\mu }>0 \) the network will classify correctly
the example with classification label \( \sigma _{\mathbf{B}\mu } \), while
if \( h\sigma _{\mathbf{B}\mu }<0 \), the classification is wrong. Thus it
gives a signal of how wrong or correct was the classification and also how stable
that classification is under changes of the weight vector. This is obviously
an important factor to take into account while incorporating the information
in a given example. The second functional structure we will concentrate on is
something that can estimate the performance or acquired experience of the network
in the implementation of the rule. This, if properly used is akin to annealing
of the learning rate or of the functional annealing in learning algorithms.
This can be implemented by using the length of the weight vector \( {\bf J}_{\mu } \).
In fig. \ref{errorj} we show a graph of \( \left\Vert {\bf J}\right\Vert  \)
as a function of the generalization error for a program with a good fitness
in the later stages of the simulation. The monotonic behavior is a typical result.
 It can be shown, at least in the thermodynamic limit, that for algorithms which
do not measure surprise their generalization error decays as \( \mu ^{-1/2} \)
and for them annealing is useless. Learning algorithms that use surprise have
a better performance \( \left( e_{g}\propto \mu ^{-1}\right)  \) and algorithms
that use both surprise and annealing by experience have an even better performance
since can have smaller coefficients of \( \mu ^{-1} \). A crude measure of
the capacity of a population of using a functional structure may be given by
the frequency that the combination of variables is found. This is admittedly
crude since the position in the program determines whether it is useful or not.
On the other hand the absence of such combination does not rule out the possibility
that some other combination is doing the job in a more cumbersome manner. In
fig. \ref{syeg} we plot the density of pairs \( h\sigma _{\mathbf{B}\mu } \)
(surprise) and the density of pairs \( \mathbf{JJ} \) (performance) in the
entire population, as functions of the number of generations. 

\noindent It is possible
to observe a fast change in the frequency of pairs of symbols related surprise
before 20 generations. \( \mathbf{JJ} \) pairs are almost immediately all but
extinguished from the population. \( h\sigma _{\mathbf{B}\mu } \) pairs are
distributed very frequently and its presence oscillates across the population
and through the generations, while \( \mathbf{JJ} \) pairs introduced by mutations
are not able to invade the population. At the time of the transition, surprise
is being correctly measured and now the appearance of \( \mathbf{JJ} \) leads
to an improvement in fitness since it leads to an estimate of the generalization
error and permits the implementation of correct annealing schedules. This successful
strategy invades the population. It is reasonable to associate the improvement
in the fitness with the emergent use of experience by the elements of the population.
Notice that injections through mutations of performance structures were non
invading before the transition. Of course this can be explained by claiming
that not every kind of annealing is beneficial but most important, before surprise
is measured correctly, no annealing scheme is useful, and therefore individuals
which could measure \( \mathbf{JJ} \) did not benefit from such knowledge. 

\begin{figure}
\epsfxsize=.98\textwidth
\epsfbox{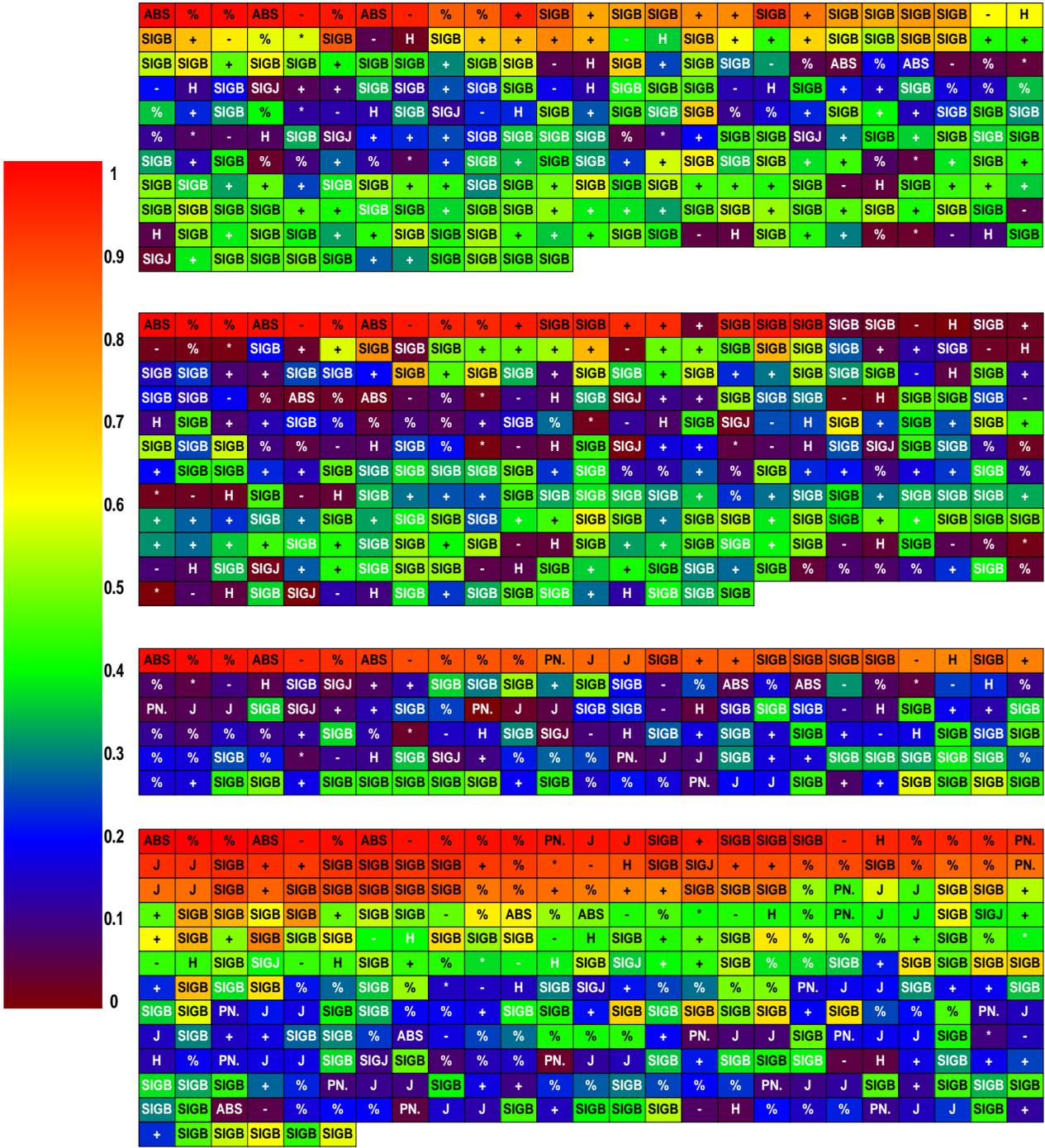}
\caption{The strings of symbols are the programs best-of-generation at generations 300,
350, 400 and 450. The colors represent the frequencies \protect\( \omega \left( s_{q}|i\right) \protect \),
according to the color scale.} 
\label{bicho0}
\end{figure}

The sequence of symbols of the BOG individual before and after the change in
the density of pairs \( \mathbf{JJ} \) mirror that increase. In fig. \ref{bicho0}
we present the most adapted individual at generations 300, 350, 400 and 450.
Just before the transition there is no pair \( \mathbf{JJ} \) \textbf{}present
in the program (the two first programs). After the transition the best individual
suffers a decrease in size and several pairs \( \mathbf{JJ} \) \textbf{}appear.
According to the color scale, red symbols are extremely frequent in the population
at that position, green symbols are just frequent at that position, and violet
are quite unlikely to be found. We can see that after the transition, the third
program, presents symbols mostly in the violet. 50 generations later there are
islands of green in the BOG. That means that the genetic character of the best
individual has invaded the population.

\begin{figure}
\epsfxsize=.48\textwidth
\epsfbox{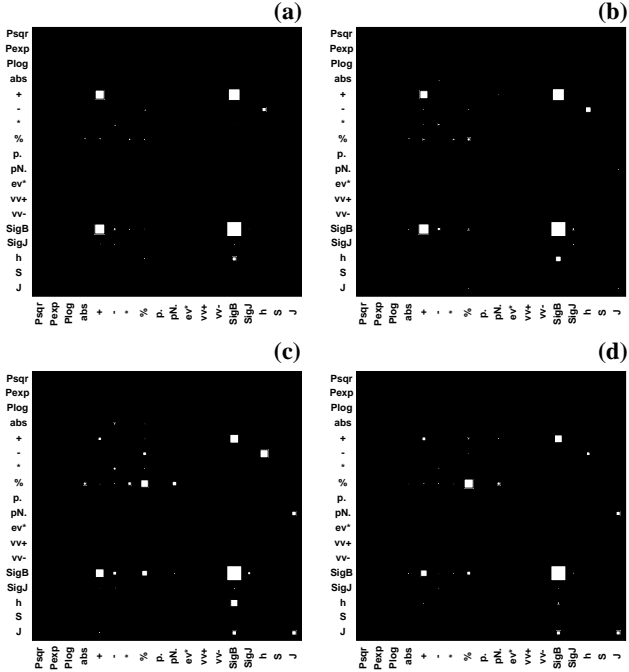}
\caption{Density of pairs for generations 300 (a), 350 (b), 400 (c), and 450 (d). In
all the cases the most frequent pair is \protect\( \sigma _{\mathbf{B}\mu }\sigma _{\mathbf{B}\mu }\protect \)
({\sf SigB SigB}). Only in the last two panels the pair \textbf{JJ} appears.}
\label{pares}
\end{figure}

A more general analysis of the density of pairs can be done with the help of
fig. \ref{pares}. In these pictures we present the relative frequencies at
which each possible pair appear in the population. The vertical axis represents
the first element of the pair, the horizontal axis the second element. The size
of the white squares represent the frequency of the pair, relative to the most
frequent pair (represented by the largest square in each picture). In panel
(a) we present the density of pairs at generation 300, (b) corresponds to generation
350, (c) to generation 400 and (d) to generation 450. In (a) and (b) there are
no pairs \( \mathbf{JJ} \). The most frequent pair is the combination \( \sigma _{\mathbf{B}\mu }\sigma _{\mathbf{B}\mu } \),
which is just a \( 1 \), but not quite since it can evolve into different directions.
After the transition, in panels (c) and (d), this pair remains the most frequent,
but important changes have happened. There are small white squares for the pair
\( \mathbf{JJ} \) \textbf{}representing the emergence of the use of experience
by the learning algorithms.

\end{multicols}\begin{multicols}{2}

\begin{figure}
\epsfxsize=.98\textwidth
\epsfbox{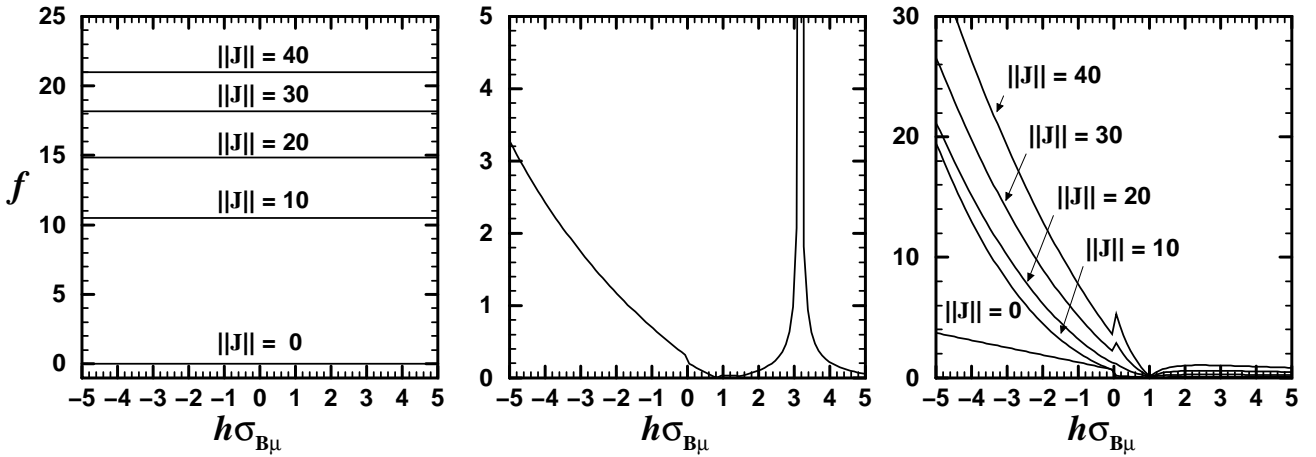}
\caption{Modulation functions. (Left) Early stage where surprise is not measured and
annealing by experience is ineffective. (Center) Intermediate stage, now surprise
is used but annealing by experience has been lost. (Right) Late stage, after
the transition, where surprise through the measurement of \protect\( h\sigma _{\mathbf{B}\mu }\protect \)
and annealing \protect\( \left\Vert {\bf J}\right\Vert \protect \) are correctly
implemented.}
\label{fmod}
\end{figure}

The modulation functions of BOG's at different stages of the evolution
can also be understood along the line of surprise-performance analysis. At earlier
stages the BOG is unable to use surprise. Although surprise functional structures
are found throughout the population, their incorrect use makes the BOG an annealed
Hebbian algorithm. It is known that annealing will not improve the Hebbian learning
and the frequency of performance functional structures decreases until it vanishes.
It will only appear in very modest ways through mutation and, repeatedly individuals
which use it becomes extinct. Later on surprise is finally well accounted for
and correctly classified examples cause typically smaller Hebbian corrections
than those incorrectly classified. At that point the correct use of surprise
potentializes the beneficial use of functional structures that measure performance.
Then a correctly annealed algorithm emerges that resembles quite closely the
modulation functions found through Bayesian or variational approaches (fig.
\ref{fmod}).
  
\section{Conclusions}

Evolutionary programming techniques provide the means to automatically design
programs which solve certain class of problems. In this paper, however we were
not interested in the final result, the problem that GP was set out to solve
has been previously analyzed from many angles and a detailed understanding of
online learning in perceptrons has been achieved. Rather we concentrated on
the dynamics of evolution and have detected dynamical changes in the behavior
of the GP solutions that we have not hesitated in dubbing dynamical transitions.
This is not a conventional phase transition associated to singularities arising
in the thermodynamic limit. A few runs failed to present the transition, maybe
because of time limitations, but it was seen in many different runs. Some features
were never reproducible but others were present in every transition. As examples
of those features that depend upon contingencies we include the number of generations
before the transition takes place, the width of the transitions (some were just
about ten generations wide, others took several tens of generations) and the
result of the GP, i.e. the program that implements the best learning algorithm.
These are mainly important from a constructive point of view when the solution
to the problem is the main concern. We tried, instead to identify robust features
which can be confidently expected to occur every time the transition takes place.
In serving such purpose we have characterized the dynamics by looking at Ph-
and G-entropies which give a picture of the distribution of phenotypic fitness
and functional or symbolic structure respectively. Large entropic fluctuations
are well described by power laws.

\end{multicols}\begin{multicols}{2}

\noindent The conformation diagram gives a bird's
eye view of the relation of the BOG and the frequency of symbols in the population
as well as its length The main robust feature can be identified once the transition
has been understood from a functional point of view, in terms of two concepts:
the surprise that newly arrived information elicits and how such information
should be taken into account based on how much experience the network has in
solving the task at hand. A temporal order can be identified in every transition.
It was never found otherwise. Performance can be useful only after surprise
is measured correctly.  

There are several possible extensions of this problem. From a biological point
of view there is a suggestive similarity with the time order in which certain
structures responsible for measuring surprise and performance have appeared.
Will this order be found in more complex artificial settings? Is this biologically
significant? Can it be extended to other functional structures? It should also
be quite interesting to further analyze phase transitions in the automatic design
of programs.

\section*{Acknowledgements}

The simulations described here were done on a cluster made possible through
the efforts of J. L. deLyra, C. E. I. Carneiro and coworkers. The cluster's
construction was partially supported by FAPESP and CNPq. JPN received financial
support from FAPESP and NC received partial support from CNPq. Discussions with
Osame Kinouchi and Mauro Copelli where important during the earlier stages of
this work.

\end{multicols}
\end{document}